# On the completeness
# of quantum computation models

Pablo Arrighi
École normale supérieure de Lyon
LIP, 46 allée d'Italie, 69008 Lyon, France
and Université de Grenoble
LIG, 220 rue de la chimie, 38400 Saint Martin d'Hères, France.
pablo.arrighi@imag.fr

Gilles Dowek
École polytechnique and INRIA,
LIX, École polytechnique, 91128 Palaiseau Cedex, France.
gilles.dowek@polytechnique.edu

**Abstract.** The notion of computability is stable (*i.e.* independent of the choice of an indexing) over infinite-dimensional vector spaces provided they have a finite "tensorial dimension". Such vector spaces with a finite tensorial dimension permit to define an absolute notion of completeness for quantum computation models and give a precise meaning to the Church-Turing thesis in the framework of quantum theory.

## 1 Introduction

In classical computing, algorithms are sometimes expressed by boolean circuits. *E.g.* the algorithm mapping the booleans $x$ and $y$ to $and(x, not(y))$ can be expressed by a circuit formed with an *and* gate and a *not* gate. A typical result about such circuit models is that a given set of gates is complete, *i.e.* that all functions from $\{0,1\}^n$ to $\{0,1\}^p$ can be expressed with a circuit with $n$ inputs and $p$ outputs. Yet, circuit models are limited because an algorithm must take as input and return as output a fixed number of boolean values, and not, for instance, an arbitrary natural number or sequence of boolean values of arbitrary length. Thus, other computation models, such as Turing machines, $\lambda$-calculus, cellular automata, ... have been designed. Such models cannot express all functions from $\mathbb{N}$ to $\mathbb{N}$, but a typical completeness result is that they can express all *computable* functions. Thus, before stating and proving such a result, we need to define a subset of the set of the functions from $\mathbb{N}$ to $\mathbb{N}$: the set of *computable* functions.
In quantum computing, both types of models exist. Most algorithms are expressed using the quantum circuit model, but more evolved models, allowing to express algorithms working with unbounded size data exist also, *e.g.* quantum

Turing machines [9,7,13], quantum $\lambda$-calculi [1,17,18,2,3], quantum cellular automata [16,6,5]. To define a notion of completeness for these quantum models of computation, we must first define a set of computable quantum functions.

Quantum functions are linear functions from a vector space to itself, and when this vector space is finite or countable (*i.e.* when it is a vector space of finite or countable dimension over a finite or countable field), the notion of computability can be transported from the natural numbers to this vector space, with the use of an *indexing*.

The robustness of the notion of computability over the natural numbers has been pointed out for long, and is embodied by the Church-Turing thesis. In contrast, it is well-known that the notion of computability over an arbitrary countable set depends on the choice of an indexing: the composition of an indexing with a non computable function yields another indexing that defines a different set of computable functions [10,8]. Thus, the robustness of the computability over the natural numbers does not directly carry through to countable vector spaces and the naive notion of completeness for such a quantum computational model is relative to the choice of an indexing.

On the other hand, a vector space is not just a set, but a set equipped with an addition and a multiplication by a scalar. More generally, sets are rarely used in isolation, but come in algebraic structures, equipped with operations. As their name suggests, these operations should be computable. Thus, in algebraic structures, we can restrict to the so called *admissible* indexings that make these operations computable [14]. In many cases, restricting to such admissible indexings makes the set of computable functions independent of the choice of the indexing. In such a case, computability is said to be *stable* over the considered algebraic structures [19].

Unfortunately, we show in this paper that, although computability is stable over finite-dimensional vector spaces, it is not stable over infinite-dimensional ones, such as those we use to define quantum functions on unbounded size data. Indeed, any non-computable permutation of the vectors of a basis changes the set of computable functions without affecting the computability of the operations. Thus, making the operations of the vector space computable is not enough to have a stable notion of computability. This result may be considered as worrying because infinite-dimensional vector spaces are rather common in physics: they are the usual sort of space in which one defines some dynamics, whether quantum or classical. As one may wish to study the ability of a physical machine or a dynamics, to compute, one would like to have a stable notion of computability on such a space, and not a notion relative to the choice of an indexing. The fact that, in quantum theory, one does not directly observe the state vector but rather the induced probability distribution over measurement outcomes does not circumvent this problem: as was pointed out [11], uncomputable amplitudes can still be read out probabilistically.

Fortunately, we show that if we add a tensor product as a primitive operation of the algebraic structure, and restrict to indexings that make also this operation computable, then computability may be stable despite the infinite dimension of the space. This theorem is obtained from a novel general result on extensions of algebraic structures which preserve stable computability.

## 2 Stable computability

We assume familiarity with the notion of a computable function over the natural numbers, as presented for instance in [15,12]. As usual, this notion is transported to finite trees labeled in a finite set through the use of an indexing. To define this indexing, we first associate an index $\ulcorner f \urcorner$ to each element $f$ of the set of labels and we define the indexing of the tree $f(t_1, \ldots, t_l)$ as $\ulcorner f(t_1, \ldots, t_l) \urcorner = \ulcorner f \urcorner; \ulcorner t_1 \urcorner; \ldots; \ulcorner t_l \urcorner; 0$ where ; is a computable one-to-one mapping from $\mathbb{N}^2$ to $\mathbb{N} \setminus \{0\}$, e.g. $n; p = (n+p)(n+p+1)/2 + n + 1$. The *indexing* of this set of trees is the partial function $i$ mapping the index $\ulcorner t \urcorner$ to the term $t$. Notice that this indexing is independent of the choice of the indexing of the labels, that is if $i_1$ and $i_2$ are two indexings of the same set of trees built on different indexings of the labels, there exists a computable function $h$ such that $i_2 = i_1 \circ h$.

When $A$ is a set of natural numbers, we say that $A$ is *effectively enumerable* if it is either empty or the image of a total computable function from $\mathbb{N}$ to $\mathbb{N}$. This is equivalent to the fact that $A$ is the image of a partial computable function, or again to the fact that it is the domain of partial computable function. We shall use the following result several times.

**Proposition 1 (Inverse).** *If $A$ and $B$ are effectively enumerable sets and $h$ is a partial computable function such that for all $x$ in $A$, $h$ is defined at $x$ and $h(x)$ is in $B$, and $h$ is surjective, i.e. for each $y$ in $B$ there exists a $x$ in $A$ such that $h(x) = y$, then $h$ has a computable right inverse, i.e. there exists a partial computable function $k$ such that for all $y$ in $B$, $k$ is defined at $y$, $h$ is defined at $k(y)$ and $h(k(y)) = y$.*

*Proof.* As the set $A$ is effectively enumerable, it is either empty, in which case $B$ also is empty and the result is trivial, or it is the image of a total recursive function $g$, we define $k_1(y)$ to be the least $z$ such that $h(g(z)) = y$ and $k(y) = g(k_1(y))$ and we have $h(k(y)) = h(g(k_1(y))) = y$.

### 2.1 Indexings

**Definition 1 (Indexing).** *An indexing of a set $E$ is a partial function $i$ from $\mathbb{N}$ to $E$, such that*

- *$dom(i)$ is effectively enumerable,*
- *$i$ is surjective,*
- *there exists a computable function eq from $dom(i) \times dom(i)$ to $\{0, 1\}$ such that $eq(x, y) = 1$ if $i(x) = i(y)$ and $eq(x, y) = 0$ otherwise.*

*An indexing of a family of sets $E_1, \ldots, E_n$ is a family of indexings $i_1, \ldots, i_n$ of the sets $E_1, \ldots, E_n$.*

Notice that this definition is slightly more general than the usual one (see for instance [19], Definition 2.2.2) as we only require $dom(i)$ to be effectively enumerable, instead of requiring it to be decidable. Requiring full decidability of $dom(i)$ is not needed in this paper and would be an obstacle to Proposition 11.

**Definition 2 (Computable function over a set equipped with an indexing).** *Let $E_1, \ldots, E_m, E_{m+1}$ be a family of sets and $i_1, \ldots, i_m, i_{m+1}$ an indexing of this family. A function from $E_1 \times \ldots \times E_m$ to $E_{m+1}$ is said to be computable relatively to this indexing if there exists a computable function $\hat{f}$ from $dom(i_1) \times \ldots \times dom(i_m)$ to $dom(i_{m+1})$ such that for all $x_1$ in $dom(i_1)$, ..., $x_m$ in $dom(i_m)$, $\hat{f}(x_1, \ldots, x_m)$ is in $dom(i_{m+1})$ and $i_{m+1}(\hat{f}(x_1, \ldots, x_m)) = f(i_1(x_1), \ldots, i_m(x_m))$.*

**Definition 3 (Admissible indexing of an algebraic structure [14]).** *An indexing of an algebraic structure $\langle E_1, \ldots, E_n, op_1, \ldots, op_p \rangle$ is an indexing of the family $E_1, \ldots, E_n$. Such an indexing is said to be* admissible *if the operations $op_1, \ldots, op_p$ are computable relatively to the indexings of their domains and co-domain.*

## 2.2 Stability

**Definition 4 (Stable computability [19]).** *Computability is said to be* stable *over a structure $\langle E_1, \ldots, E_n, op_1, \ldots, op_p \rangle$ if there exists an admissible indexing $i_1, \ldots, i_n$ of this structure and for all admissible indexings $j_1, \ldots, j_n$ and $j'_1, \ldots, j'_n$ of this structure, there exist computable functions $h_1$ from $dom(j'_1)$ to $dom(j_1)$, ..., $h_n$ from $dom(j'_n)$ to $dom(j_n)$ such that $j'_1 = j_1 \circ h_1$, ..., $j'_n = j_n \circ h_n$.*

We now want to prove that the notion of computable function over a structure with stable computability is independent of the choice of the indexing. To state this in full generality, we consider a structure $\langle E_1, \ldots, E_n, op_1, \ldots, op_p \rangle$ with stable computability, and a family of sets $F_1, \ldots, F_m, F_{m+1}$ some of which are among $E_1, \ldots, E_n$, in which case various admissible indexings can be considered for these sets. We prove that the set of computable functions from $F_1, \ldots, F_m$ to $F_{m+1}$ is independent of these choices of admissible indexings.

**Proposition 2.** *Let $\langle E_1, \ldots, E_n, op_1, \ldots, op_p \rangle$ be a structure with stable computability. Then, for all families of sets $F_1, \ldots, F_{m+1}$ and indexings $i_1, \ldots, i_{m+1}$ and $i'_1, \ldots, i'_{m+1}$ of this family, such that for all $k$ either $i_k = i'_k$ or $F_k$ is some $E_l$, and $i_k$ and $i'_k$ are two indexings of $E_l$ extracted from two admissible indexings of the structure $\langle E_1, \ldots, E_n, op_1, \ldots, op_p \rangle$, the set of computable functions from $F_1, \ldots, F_m$ to $F_{m+1}$ equipped with $i_1, \ldots, i_{m+1}$ and $i'_1, \ldots, i'_{m+1}$ is the same.*

*Proof.* There exist computable functions $h_1, \ldots, h_m, h_{m+1}$ such that $i'_1 = i_1 \circ h_1$, ..., $i'_m = i_m \circ h_m$, $i'_{m+1} = i_{m+1} \circ h_{m+1}$. Let $h'_1, \ldots, h'_m, h'_{m+1}$ be right inverses of $h_1, \ldots, h_m, h_{m+1}$.

If there exists a computable function $\hat{f}$ such that $i_{m+1}(\hat{f}(x_1,\ldots,x_m)) = f(i_1(x_1),\ldots,i_m(x_m))$, the function $\tilde{f}$ mapping $x_1,\ldots,x_m$ to $h'_{m+1}(\hat{f}(h_1(x_1),\ldots,h_m(x_m)))$ is computable and $i'_{m+1}(\tilde{f}(x_1,\ldots,x_m)) = f(i'_1(x_1),\ldots,i'_m(x_m))$. Conversely, if there exists a function $\tilde{f}$ such that $i'_{m+1}(\tilde{f}(x_1,\ldots,x_m)) = f(i'_1(x_1),\ldots,i'_m(x_m))$, the function $\hat{f}$ mapping $x_1,\ldots,x_m$ to $h_{m+1}(\tilde{f}(h'_1(x_1),\ldots,h'_m(x_m)))$ is computable and $i_{m+1}(\hat{f}(x_1,\ldots,x_m)) = f(i_1(x_1),\ldots,i_m(x_m))$.

We can also prove the converse.

**Proposition 3.** *Let $\langle E_1,\ldots,E_n, op_1,\ldots,op_p\rangle$ be a structure. If for all families of sets $F_1,\ldots,F_{m+1}$ and indexings $i_1,\ldots,i_{m+1}$ and $i'_1,\ldots,i'_{m+1}$ of this family, such that for all $k$ either $i_k = i'_k$ or $F_k$ is some $E_l$, $i_k$ and $i'_k$ are extracted from two admissible indexings of the structure $\langle E_1,\ldots,E_n, op_1,\ldots,op_p\rangle$ the set of computable functions from $F_1,\ldots,F_m$ to $F_{m+1}$ equipped with $i_1,\ldots,i_{m+1}$ and with $i'_1,\ldots,i'_{m+1}$ is the same. Then, computability is stable over the structure $\langle E_1,\ldots,E_n, op_1,\ldots,op_p\rangle$.*

*Proof.* Let $j_1,\ldots,j_n$ and $j'_1,\ldots,j'_n$ be two admissible indexings of $\langle E_1,\ldots,E_n, op_1,\ldots,op_p\rangle$. The identity from $E_k$ to $E_k$ is computable relatively to $j_k, j_k$. Our hypothesis tells us that it also computable relatively to $j'_k, j_k$, *i.e.* that there exists a computable function $h_k$, from $dom(j'_k)$ to $dom(j_k)$, such that for all $x$, $j_k(h_k(x)) = j'_k(x)$ *i.e.* $j_k \circ h_k = j'_k$.

## 3 Relative finite generation

We now want to prove that computability is stable over finitely generated structures. Intuitively, a structure is finitely generated if all its elements can be constructed with the operations of the structure from a finite number of elements $a_0,\ldots,a_{d-1}$. For instance, the structure $\langle \mathbb{N}, S\rangle$ is finitely generated because all natural numbers can be generated with the successor function from the number 0. This property can also be stated as the fact that for each element $b$ of the structure there exists a term $t$, *i.e.* a finite tree labeled with the elements $a_0,\ldots,a_{d-1}$ and the operations of the structure, such that the denotation $[\![t]\!]$ of $t$ is $b$. Yet, in this definition, we must take into account two other elements.

First, in order to prove that computability is stable, we must add the condition that equality of denotations is decidable on terms. This leads us to sometimes consider not the full set of terms $\mathcal{T}$ generated by the operations of the structure, but just an effectively enumerable subset $T$ of this set, which is chosen large enough to contain a term denoting each element of the structure. This is in order to avoid situations where the equality of denotations may be undecidable, or more at least more difficult to prove decidable, on the full set of terms.

Secondly, when a structure has several domains, such as a vector space that has a domain for scalars and a domain for vectors, we want, in some cases, to assume first that some substructure, *e.g.* the field of the scalars, has stable computability, and still be able to state something alike finite generation for the other domains, *e.g.* the domain of the vectors. Thus, we shall consider structures $\langle E_1, \ldots, E_m, E_{m+1}, \ldots, E_{m+n}, op_1, \ldots, op_p, op_{p+1}, \ldots, op_{p+q} \rangle$ with two kinds of domains, assume that the structure $\langle E_1, \ldots, E_m, op_1, \ldots, op_p \rangle$ has stable computability, and consider terms that generate the elements of $E_{m+1}, \ldots, E_{m+n}$ with the operations $op_{p+1}, \ldots, op_{p+q}$. In these terms, the elements of $E_1, \ldots, E_m$ are expressed by their index which is itself expressed using the symbols 0 and $S$, in order to keep the language finite. The case $m = 0$ is the usual definition of finite generation, whereas in the case $m > 0$ we say that the algebraic structure $\langle E_{m+1}, \ldots, E_{m+n}, op_{p+1}, \ldots, op_{p+q} \rangle$ is finitely generated relatively to $\langle E_1, \ldots, E_m, op_1, \ldots, op_p \rangle$, which in turn may or may not be finitely generated.

**Definition 5 (Terms, denotation).** *Let $E_1, \ldots, E_m$ be sets equipped with indexings $i_1, \ldots, i_m$, $E_{m+1}, \ldots, E_{m+n}$ be sets, $A = \{a_0, \ldots, a_{d-1}\}$ be a finite set of elements of $E_{m+1}, \ldots, E_{m+n}$, and $op_{p+1}, \ldots, op_{p+q}$ be operations whose arguments are in $E_1, \ldots, E_m, E_{m+1}, \ldots, E_{m+n}$ but whose values are in $E_{m+1}, \ldots, E_{m+n}$. The set $\mathcal{T}_k$ of terms of sort $E_k$ is inductively defined as follows*

- *if the natural number $x$ is an element of $dom(i_k)$ (for $k$ in $1, \ldots, m$), then $S^x(0)$ is a term of sort $E_k$,*
- *if $a$ is an element of $A$ and $E_k$, then $a$ is a term of sort $E_k$,*
- *if $t_1$ is a term of sort $E_{k_1}$, ..., $t_l$ is an term of sort $E_{k_l}$ and $op$ is one of the functions from $op_{p+1}, \ldots, op_{p+q}$ from $E_{k_1} \times \ldots \times E_{k_l}$ to $E_{k_{l+1}}$ then $op(t_1, \ldots, t_l)$ is a term of sort $E_{k_{l+1}}$.*

*The denotation of a term of sort $E_k$ is the element of $E_k$ defined as follows*

- $[\![S^x(0)]\!]_{E_k} = i_k(x)$,
- $[\![a]\!]_{E_k} = a$,
- $[\![op(t_1, \ldots, t_l)]\!]_{E_{k_{l+1}}} = op([\![t_1]\!]_{E_{k_1}}, \ldots, [\![t_l]\!]_{E_{k_l}})$.

*As a term is a tree labeled in the finite set $a_0, \ldots, a_{d-1}, op_{p+1}, \ldots, op_{p+q}, 0, S$, we can associate an index $\ulcorner t \urcorner$ to each term $t$ in a canonical way.*

**Definition 6 (Finite generative set).** *Let $E_1, \ldots, E_m$ be a family of sets equipped with indexings $i_1, \ldots, i_m$ and let $\langle E_{m+1}, \ldots, E_{m+n}, op_{p+1}, \ldots, op_{p+q} \rangle$ be a structure. A finite set $A$ of elements of the sets $E_{m+1}, \ldots, E_{m+n}$ is said to be a finite generative set of the structure $\langle E_{m+1}, \ldots, E_{m+n}, op_{p+1}, \ldots, op_{p+q} \rangle$ relatively to $E_1, \ldots, E_m, i_1, \ldots, i_m$, if there exist effectively enumerable subsets $T_{m+1}, \ldots, T_{m+n}$ of the sets $\mathcal{T}_{m+1}, \ldots, \mathcal{T}_{m+n}$ of terms of sort $E_{m+1}, \ldots, E_{m+n}$, such that for each element $b$ of a set $E_k$ there exists a term $t$ in $T_k$ such that $b = [\![t]\!]_{E_k}$ and, for each $k$, there exists a computable function $eq_k$ such that for all $t$ and $u$ in $T_k$, $eq_k(\ulcorner t \urcorner, \ulcorner u \urcorner) = 1$ if $[\![t]\!] = [\![u]\!]$ and $eq_k(\ulcorner t \urcorner, \ulcorner u \urcorner) = 0$ otherwise.*

**Definition 7 (Finitely generated).** *Let $E_1, \ldots, E_m$ be a family of sets equipped with indexings $i_1, \ldots, i_m$. The structure $\langle E_{m+1}, \ldots, E_{m+n}, op_{p+1}, \ldots, op_{p+q} \rangle$ is said to be finitely generated relatively to $E_1, \ldots, E_m, i_1, \ldots, i_m$, if it has a finite generative set relatively to $E_1, \ldots, E_m, i_1, \ldots, i_m$.*

*Remark 1.* This notion of finite generation generalizes to arbitrary structures the notion of finite-dimensional vector space. More generally, we can define the *dimension* of a structure as the minimal cardinal of a finite generative set of this structure.

**Theorem 1.** *Let $\langle E_1, \ldots, E_m, op_1, \ldots, op_p \rangle$ be a structure with stable computability and $s_1, \ldots, s_m$ be an admissible indexing of this structure. Then, if the structure $\langle E_{m+1}, \ldots, E_{m+n}, op_{p+1}, \ldots, op_{p+q} \rangle$ is finitely generated relatively to $E_1, \ldots, E_m, s_1, \ldots, s_m$, then computability is stable over the structure $\langle E_1, \ldots, E_m, E_{m+1}, \ldots, E_{m+n}, op_1, \ldots, op_p, op_{p+1}, \ldots, op_{p+q} \rangle$.*

*Proof.* We first prove that there exists an admissible indexing of this structure. If $k$ is an element of $1, \ldots, m$, we let $i_k$ be the function mapping the index $\ulcorner S^n(0) \urcorner$ to $[\![S^n(0)]\!]_{E_k} = s_k(n)$. By construction, the domain of $i_k$ is effectively enumerable, $i_k$ is surjective, and equality is decidable. If $k$ is an element of $m+1, \ldots, m+n$, we let $i_k$ be the function mapping the index $\ulcorner t \urcorner$ of the term $t$ of $T_k$ to $[\![t]\!]_{E_k}$. The domain of $i_k$ is the set of indices of elements of $T_k$, thus it is effectively enumerable, for every element $b$ of $E_k$, there exists a term $t$ of $T_k$ such that $b = [\![t]\!]_{E_k}$, thus the function $i_k$ is surjective. Finally, equality is decidable by hypothesis. Thus, $i_1, \ldots, i_{m+n}$ is an indexing of the structure $\langle E_1, \ldots, E_{m+n}, op_1, \ldots, op_{p+q} \rangle$. Let us prove it is admissible.

Let $op$ be one of the functions $op_1, \ldots, op_p$. As $s_1, \ldots, s_m$ is an admissible indexing of $\langle E_1, \ldots, E_m, op_1, \ldots, op_p \rangle$, there exists a function $\hat{op}$ such that $s_{l+1}(\hat{op}(x_1, \ldots, x_l)) = op(s_1(x_1), \ldots, s_l(x_l))$. We define the function $\tilde{op}$ as the function mapping $\ulcorner S^{x_1}(0) \urcorner, \ldots, \ulcorner S^{x_l}(0) \urcorner$ to $\ulcorner S^{\hat{op}(x_1, \ldots, x_l)}(0) \urcorner$. Let $y_1$ be an element of $dom(i_{k_1})$, $\ldots$, $y_l$ be an element of $dom(i_{k_l})$, there exists $x_1, \ldots, x_l$ such that $y_1 = \ulcorner S^{x_1}(0) \urcorner, \ldots, y_l = \ulcorner S^{x_l}(0) \urcorner$ and we have $i_{k_{l+1}}(\tilde{op}(y_1, \ldots, y_l)) = i_{k_{l+1}}(\tilde{op}(\ulcorner S^{x_1}(0) \urcorner, \ldots, \ulcorner S^{x_l}(0) \urcorner)) = i_{k_{l+1}}(\ulcorner S^{\hat{op}(x_1, \ldots, x_l)}(0) \urcorner) = s_{k_{l+1}}(\hat{op}(x_1, \ldots, x_l)) = op(s_{k_1}(x_1), \ldots, s_{k_l}(x_l)) = op(i_{k_1}(\ulcorner S^{x_1}(0) \urcorner), \ldots, i_{k_l}(\ulcorner S^{x_l}(0) \urcorner)) = op(i_{k_1}(y_1), \ldots, i_{k_l}(y_l))$. Thus, the operations $op_1, \ldots, op_p$ are computable. Now, if $op$ is one of the functions $op_{p+1}, \ldots, op_{p+q}$, we let $\tilde{op}$ be the computable function mapping $x_1, \ldots, x_l$ to $\ulcorner op \urcorner; x_1; \ldots; x_l; 0$. Notice that this function maps the indices $\ulcorner t_1 \urcorner, \ldots, \ulcorner t_l \urcorner$ to the index $\ulcorner op(t_1, \ldots, t_l) \urcorner$. Let $x_1, \ldots, x_l$ be elements of $dom(i_{k_1}), \ldots, dom(i_{k_l})$. There exists terms $t_1, \ldots, t_l$ such that $x_1 = \ulcorner t_1 \urcorner, \ldots, x_l = \ulcorner t_l \urcorner$. We have $i_{k_{l+1}}(\tilde{op}(x_1, \ldots, x_l)) = i_{k_{l+1}}(\tilde{op}(\ulcorner t_1 \urcorner, \ldots, \ulcorner t_l \urcorner)) = i_{k_{l+1}}(\ulcorner op(t_1, \ldots, t_l) \urcorner) = [\![op(t_1, \ldots, t_l)]\!] = op([\![t_1]\!], \ldots, [\![t_l]\!]) = op(i_{k_1}(\ulcorner t_1 \urcorner), \ldots, i_{k_l}(\ulcorner t_l \urcorner)) = op(i_{k_1}(x_1), \ldots, i_{k_l}(x_l))$. Thus, the operations $op_{p+1}, \ldots, op_{p+q}$ are computable too. Hence, the indexing $i_1, \ldots, i_{m+n}$ is admissible.

If $j_1, \ldots, j_{m+n}$ is an arbitrary admissible indexing of the structure $\langle E_1, \ldots, E_{m+n}, op_1, \ldots, op_{p+q} \rangle$ with computable functions $\tilde{op}_1, \ldots, \tilde{op}_{p+q}$. The indexing $j_1, \ldots, j_m$ is an admissible indexing of the structure $\langle E_1, \ldots, E_m, op_1, \ldots, op_p \rangle$. Thus, there exists computable functions $g_1, \ldots, g_m$ such that $s_k = j_k \circ g_k$. For $k$ in $1, \ldots, m$, we define the computable function $h_k$ as the function mapping $\ulcorner S^x(0) \urcorner$ to $g_k(x)$. For $k$ in $m+1, \ldots, m+n$, we define the computable functions $h_k$ by induction:

- the value of $h_k$ on the index of a constant $a$ of $A$ is the least $x$ such that $j_k(x) = a$,
- the value of $h_k$ on the index $\ulcorner op(t_1, \ldots, t_l) \urcorner$ is $\tilde{op}(h_{k_1}(\ulcorner t_1 \urcorner), \ldots, h_{k_l}(\ulcorner t_l \urcorner))$.

Then, we prove by induction over term structure that, for all terms $t$ of sort $E_k$, $j_k(h_k(\ulcorner t \urcorner)) = i_k(\ulcorner t \urcorner)$. If $t$ has the form $S^x(0)$ then we have $j_k(h_k(\ulcorner t \urcorner)) = j_k(h_k(\ulcorner S^x(0) \urcorner)) = j_k(g_k(x)) = s_k(x) = i_k(\ulcorner S^x(0) \urcorner) = i_k(t)$. If $t$ is a constant $a$, we have $j_k(h_k(\ulcorner a \urcorner)) = a = [\![a]\!]_{E_k} = i_k(\ulcorner a \urcorner)$ and if $t$ has the form $op(t_1, \ldots, t_l)$, for some operation $op$, we have $j_{k_{l+1}}(h_{k_{l+1}}(\ulcorner op(t_1, \ldots, t_l) \urcorner)) = j_{k_{l+1}}(\tilde{op}(h_{k_1}(\ulcorner t_1 \urcorner), \ldots, h_{k_l}(\ulcorner t_l \urcorner))) = op(j_{k_1}(h_{k_1}(\ulcorner t_1 \urcorner)), \ldots, j_{k_l}(h_{k_l}(\ulcorner t_l \urcorner))) = op(i_{k_1}(\ulcorner t_1 \urcorner), \ldots, i_{k_l}(\ulcorner t_l \urcorner)) = op([\![t_1]\!], \ldots, [\![t_l]\!]) = [\![op(t_1, \ldots, t_l)]\!] = i_{k_{l+1}}(\ulcorner op(t_1, \ldots, t_l) \urcorner)$. Thus, for all elements of the domain of $i_k$, $j_k(h_k(x)) = i_k(x)$, i.e. $j_k \circ h_k = i_k$.

Finally, if $j_1, \ldots, j_{m+n}$ and $j'_1, \ldots, j'_{m+n}$ are two indexings of the structure $\langle E_1, \ldots, E_{m+n}, op_1, \ldots, op_{p+q} \rangle$. There exists functions $h_1, \ldots, h_{m+n}$ and $h'_1, \ldots, h'_{m+n}$ such that for all $k$, $j_k \circ h_k = i_k$ and $j'_k \circ h'_k = i_k$. Thus, for all $k$, $j_k \circ h_k = j'_k \circ h'_k$. Let $h''_1, \ldots, h''_{m+n}$ be right inverses of $h'_1, \ldots, h'_{m+n}$. We have $j'_k = j_k \circ (h_k \circ h''_k)$. Hence, computability is stable over the structure $\langle E_1, \ldots, E_{m+n}, op_1, \ldots, op_{p+q} \rangle$.

As corollaries, we get stable computability for well-known cases.

**Proposition 4 (Natural numbers).** *Computability is stable over the structures $\langle \mathbb{N}, S \rangle$ and $\langle \mathbb{N}, + \rangle$.*

*Proof.* Consider the set of terms in the language $0, S$. Each natural number is denoted by a term. Moreover, as each natural number is denoted by a unique term, equality of denotations is trivial. Thus, the structure $\langle \mathbb{N}, S \rangle$ is finitely generated and computability is stable over this structure.

Consider the set of terms in the language $0, 1, +$ and its subsets of terms of the form $1 + (1 + \ldots + (1 + 0) \ldots)$. Each natural number is denoted by a term. Moreover, as each natural number is denoted by a unique term, equality of denotations is trivial. Thus, the structure $\langle \mathbb{N}, + \rangle$ is finitely generated and computability is stable over this structure.

*Remark 2.* For the structure $\langle \mathbb{N}, + \rangle$ we could also have considered all the terms in the language $0, 1, +$, in which case we would have had to provide an algorithm to test the equality of denotations of two such terms.

**Proposition 5 (Rational numbers).** *Computability is stable over the structure $\langle \mathbb{Q}, +, -, \times, / \rangle$.*

*Proof.* The structure $\langle \mathbb{Q}, +, -, \times, / \rangle$ is finitely generated. Consider all terms of the form $(p - q)/(1 + r)$ where $p$, $q$ and $r$ have the form $1 + (1 + \ldots + (1 + 0) \ldots)$, and either $p$ or $q$ is 0, if both are then $r = 0$ also, and $p - q$ and $1 + r$ are relatively prime otherwise. Each rational number is denoted by a term. Moreover, as each rational number is denoted by a unique term, equality of denotations is trivial. Thus, the structure $\langle \mathbb{Q}, +, -, \times, / \rangle$ is finitely generated and computability is stable over this structure.

**Proposition 6.** *Computability is stable over the structure $\langle \mathbb{Q}, +, \times \rangle$.*

*Proof.* We prove that the indexings admissible for $\langle \mathbb{Q}, +, \times \rangle$ and $\langle \mathbb{Q}, +, -, \times, / \rangle$ are the same. If the indexing $i$ is admissible for $\langle \mathbb{Q}, +, -, \times, / \rangle$ then it is obviously admissible for $\langle \mathbb{Q}, +, \times \rangle$. Conversely, if $i$ is admissible for $\langle \mathbb{Q}, +, \times \rangle$, then there exist computable functions $\hat{+}$ and $\hat{\times}$, from $\mathbb{N}^2$ to $\mathbb{N}$, such that $i(x \mathbin{\hat{+}} y) = i(x) + i(y)$ and $i(x \mathbin{\hat{\times}} y) = i(x) \times i(y)$. The set $dom(i)$ is non empty and recursively enumerable, thus it is the image of a computable function $g$. Let $n$ and $p$ be two natural numbers, we define $n \mathbin{\hat{-}} p = g(z)$ where $z$ is the least natural number such that $p \mathbin{\hat{+}} g(z) = n$. We have $i(x \mathbin{\hat{-}} y) = i(x) - i(y)$. We build the function $\hat{/}$ in a similar way. Thus, $i$ is admissible for $\langle \mathbb{Q}, +, -, \times, / \rangle$.

*Remark 3.* Computability is stable over the structure $\langle \mathbb{Q}, +, \times \rangle$, but this structure is not finitely generated. Indeed, consider a finite number of rational numbers $a_0 = p_0/q_0, \ldots, a_{d-1} = p_{d-1}/q_{d-1}$ and call $q$ a common multiple of the denominators $q_0, \ldots, q_{d-1}$. All numbers that are the denotation of a term in the language $a_0, \ldots, a_{d-1}, +, \times$ have the form $p/q^k$ for some $p$ and $k$. The numbers $q+1$ and $q$ are relatively prime, thus $q+1$ is not a divisor of $q^k$ for any $k$, the number $1/(q+1)$ does not have the form $p/q^k$, and it cannot be expressed by a term. Thus, finite generation is a sufficient condition for stable computability, but it is not a necessary one.

The stability of computability over the structure $\langle \mathbb{Q}, +, \times \rangle$ can be explained by the fact that, although subtraction and division are not operations of the structure, they can be effectively defined from these operations. Such a structure is said to be *effectively generated*.

There are several ways to define this notion. For instance, [20] proposes a definition based on a simple imperative programming language with the operations of the algebraic structure as primitive. A more abstract definition is the existence of a language, that needs not be the language of terms built with operations on the structure, but may, for instance, be a programming language, and a denotation function associating an element of the algebraic structure to each expression of the language, verifying the following properties:

- the set of well-formed expressions is effectively enumerable,
- each element is the denotation of some expression,
- equality of denotations is decidable,
- an expression denoting $op(a_1, \ldots, a_n)$ can be built from ones denoting $a_1, \ldots, a_n$,
- for any admissible indexing, an index of $a$ can be computed from an expression denoting $a$.

Replacing expressions by their indices, we get this way exactly the definition of stable computability. Thus, stability of computability and effective generation are trivially equivalent in this approach.

## 4 Vector spaces

**Proposition 7 (Vector spaces).** *If computability is stable over the field $\langle K, +, \times \rangle$ and the vector space $\langle K, E, +, \times, +, . \rangle$ is finite-dimensional, then computability is stable over $\langle K, E, +, \times, +, . \rangle$.*

*Proof.* Let $s$ be an indexing of the field $\langle K, +, \times \rangle$ and $e_0, \ldots, e_{d-1}$ be a basis of $E$. Consider the set of terms of the form $(S^{\lambda_0}(0).e_0 + (S^{\lambda_1}(0).e_1 + (\ldots S^{\lambda_{d-1}}(0).e_{d-1})))$ where $\lambda_0, \ldots, \lambda_{d-1}$ are in $dom(s)$. The denotation of such a term is the vector $s(\lambda_0).e_0 + s(\lambda_1).e_1 + \ldots + s(\lambda_{d-1}).e_{d-1}$. As the function $s$ is surjective and $e_0, \ldots, e_{d-1}$ is a basis of $E$, each element of $E$ is the denotation of such a term. Moreover, equality of denotations can be decided on scalars, it can be decided on such terms. Thus, the structure $\langle E, +, . \rangle$ is finitely generated relatively to $\langle K, +, \times \rangle$ and computability is stable over $\langle K, E, +, \times, +, . \rangle$.

**Proposition 8.** *Computability is not stable over the structure $\langle \mathbb{N} \rangle$.*

*Proof.* Let $f$ be a non computable one-to-one function from $\mathbb{N}$ to $\mathbb{N}$, for instance the function mapping $2n$ to $2n+1$ and $2n+1$ to $2n$ if $n \in U$ and $2n$ to $2n$ and $2n+1$ to $2n+1$ otherwise, where $U$ is any undecidable set.

The function $f$ and the identity are both admissible indexings of the structure $\langle \mathbb{N} \rangle$. If computability were stable over this structure, then there would exist a computable function $h$ such that $f = id \circ h$. Thus, $f = h$ would be computable, which is contradictory.

**Proposition 9.** *Computability is stable over the vector space $\langle K, E, +, \times, +, . \rangle$ if and only if it is stable over the field $\langle K, +, \times \rangle$ and $\langle K, E, +, \times, +, . \rangle$ is a finite-dimensional vector space.*

*Proof.* If the dimension of $E$ is not countable, then the set $E$ itself is not countable, and hence computability is not stable over this structure.

If the dimension of $E$ is finite or countable and computability is not stable over the field $\langle K, +, \times \rangle$, then it is not stable over the vector space $\langle K, E, +, \times, +, . \rangle$ either. Indeed, if $s$ and $s'$ are two non equivalent indexings of $\langle K, +, \times \rangle$, and $e_0, e_1, \ldots$ is a finite or countable basis of $E$, then, we let $i$ be the function mapping $\ulcorner (S^{\lambda_0}(0).e_0 + (S^{\lambda_1}(0).e_1 + (\ldots S^{\lambda_n}(0).e_n)))\urcorner$ where $\lambda_0, \lambda_1, \ldots, \lambda_n$ are in $dom(s)$ to $s(\lambda_0).e_0 + s(\lambda_1).e_1 + \ldots + s(\lambda_n).e_n$ and $i'$ be the function mapping $\ulcorner (S^{\lambda_0}(0).e_0 + (S^{\lambda_1}(0).e_1 + (\ldots S^{\lambda_n}(0).e_n)))\urcorner$ where $\lambda_0, \lambda_1, \ldots, \lambda_n$ are in $dom(s')$ to $s'(\lambda_0).e_0 + s'(\lambda_1).e_1 + \ldots + s'(\lambda_n).e_n$, and $s, i$ and $s', i'$ are two non equivalent indexings of $\langle K, E, +, \times, +, . \rangle$.

We have proved in Proposition 7 that if computability is stable over the field $\langle K, +, \times \rangle$ and the vector space $\langle K, E, +, \times, +, . \rangle$ is finite-dimensional, then computability is stable over this space. Thus, all that remains to be proved is that if computability is stable over the field $\langle K, +, \times \rangle$ and the vector space $\langle K, E, +, \times, +, . \rangle$ has a countably infinite dimension, then computability is not stable over this vector space.

Let $s$ be an indexing of $\langle K, +, \times \rangle$, $e_0, e_1, \ldots$ be a basis of $E$, and $i$ be the function mapping $\ulcorner (S^{\lambda_0}(0).e_0 + (S^{\lambda_1}(0).e_1 + (\ldots S^{\lambda_n}(0).e_n)))\urcorner$ where $\lambda_0, \lambda_1, \ldots \lambda_n$ are in $dom(s)$ to $s(\lambda_0).e_0+s(\lambda_1).e_1+\ldots+s(\lambda_n).e_n$. As the function $s$ is surjective and $e_0, e_1, \ldots$ is a basis, the function $i$ is surjective. As there exist computable functions $\hat{+}$ and $\hat{\times}$ and $eq$ on scalars, we can build computable functions $\hat{+}$ and $\hat{.}$ and $eq$ on such terms. Thus, the function $i$ is an admissible indexing of $E$.

Then, let $f$ be a non computable one-to-one function from $\mathbb{N}$ to $\mathbb{N}$. Let $\phi$ be the one-to-one linear function from $E$ to $E$ mapping the basis vector $e_p$ to $e_{f(p)}$ for all $p$. As $\phi$ is a one-to-one mapping between $E$ and itself, $\phi \circ i$ is a surjection from $dom(i)$ to $E$. As $\phi$ is linear, $\phi(i(u \hat{+} v)) = \phi(i(u)+i(v)) = \phi(i(u))+\phi(i(v))$ and $\phi(i(\lambda \hat{.} u)) = \phi(s(\lambda).i(u)) = s(\lambda).\phi(i(u))$. Thus, $\phi \circ i$ is an admissible indexing of $E$. And if computability were stable over $\langle K, E, +, \times, +, . \rangle$, there would exists a computable function $g$ from $dom(i)$ to $dom(i)$ such that $\phi \circ i = i \circ g$.

Let $z$ be any number such that $s(z)$ is the scalar $0$ and $u$ be any number such that $s(u)$ is the scalar $1$. Let $B$ be the computable function, from $\mathbb{N}$ to $\mathbb{N}$, mapping $p$ to $\ulcorner S^z(0).e_0 + \ldots + S^z(0).e_{p-1} + S^u(0).e_p \urcorner$. We have $i(B(p)) = e_p$. Let $C$ be the partial computable function from $\mathbb{N}$ to $\mathbb{N}$ mapping the index $\ulcorner (S^{\lambda_0}(0).e_0+(S^{\lambda_1}(0).e_1+(\ldots S^{\lambda_n}(0).e_n))) \urcorner$ to the least $p$ such that $eq(\lambda_p, u) = 1$. If $i(x) = e_p$ then $C(x) = p$. Let $h$ be the computable function $C \circ g \circ B$. Let $p$ be an arbitrary natural number. We have $i(g(B(p))) = \phi(i(B(p))) = \phi(e_p) = e_{f(p)}$. Thus, $h(p) = C(g(B(p))) = f(p)$. Thus, $f = h$ would be computable, which is contradictory.

## 5 Field extensions

Using Proposition 7, we get that if computability is stable over the field $\langle K, +, \times \rangle$ and the field $\langle L, +, \times \rangle$ is an extension of finite degree of this field, then computability is stable over the structure $\langle K, L, +, \times, +, . \rangle$ where . is the product of an element of $L$ by an element of $K$.

We can easily prove that computability is also stable over the structure $\langle K, L, +, \times, +, ., \times \rangle$ where the multiplication of $L$ is added.

**Proposition 10 (Field extension of finite degree).** *If computability is stable over the field $\langle K, +, \times \rangle$, and the field $\langle L, +, \times \rangle$ is an extension of finite degree of $\langle K, +, \times \rangle$, then computability is stable over the structure $\langle K, L, +, \times, +, ., \times \rangle$.*

*Proof.* Let $s$ be an indexing of the field $\langle K, +, \times \rangle$, $e_0, \ldots, e_{d-1}$ be a basis of $L$, and $i$ be the function mapping $\ulcorner (S^{\lambda_0}(0).e_0 + (S^{\lambda_1}(0).e_1 + (\ldots S^{\lambda_{d-1}}(0).e_{d-1}))) \urcorner$, where $\lambda_0, \ldots, \lambda_{d-1}$ are in $dom(s)$, to $s(\lambda_0).e_0 + s(\lambda_1).e_1 + \ldots + s(\lambda_{d-1}).e_{d-1}$. As the function $s$ is surjective and $e_0, \ldots e_{d-1}$ is a basis of $E$, the function $i$ is surjective. As there exist computable functions $\hat{+}$ and $\hat{\times}$ and $eq$ on elements of $K$, we can build computable functions $\hat{+}$ and $\hat{.}$ and $eq$ on such terms. Thus, $s, i$, is an admissible indexing of the structure $\langle K, L, +, \times, +, . \rangle$.

Let us prove that it is also an admissible indexing of the structure $\langle K, L, +, \times, +, ., \times \rangle$. The elements $e_p \times e_q$ of $L$ can be written in a unique way $\Sigma_r m_{p,q,r} e_r$. Let $\mu_{p,q,r}$ be natural numbers such that $s(\mu_{p,q,r}) = m_{p,q,r}$. Let $\hat{\times}_L$ be the computable function mapping $\ulcorner (S^{\lambda_0}(0).e_0 + (S^{\lambda_1}(0).e_1 + (\ldots S^{\lambda_{d-1}}(0).e_{d-1}))) \urcorner$ and $\ulcorner (S^{\lambda'_0}(0).e_0 + (S^{\lambda'_1}(0).e_1 + (\ldots S^{\lambda'_{d-1}}(0).e_{d-1}))) \urcorner$ to $\ulcorner (S^{\lambda''_0}(0).e_0 + (S^{\lambda''_1}(0).e_1 +$ $(\ldots S^{\lambda''_{d-1}}(0).e_{d-1})))^\rceil$ where $\lambda''_r = \hat{\Sigma}_{p,q} \lambda_p \hat{\times}_K \lambda'_q \hat{\times}_K \mu_{p,q,r}$. It is routine to check that if $x$ and $y$ are in the $dom(i)$, then $i(x \hat{\times}_L y) = i(x) \times i(y)$. Thus, $s, i$ is an admissible indexing of the structure $\langle K, L, +, \times, +, ., \times \rangle$ as well.

Finally, if $s, j$ and $s', j'$ are two admissible indexings of $\langle K, L, +, \times, +, ., \times \rangle$, then they are obviously admissible indexings of $\langle K, L, +, \times, +, . \rangle$ as well, thus there exists computable functions $h$ and $k$ such that $s' = s \circ h$ and $j' = j \circ k$.

Now, we would like to prove that computability is stable over the field $\langle L, +, \times \rangle$. But, proving this result seems to require an extra hypothesis, that $K$ is an effectively enumerable subset of $L$, i.e. that if $i$ is an admissible indexing of $L$, we can enumerate the indices of the elements of $K$. As this is always the case when $K$ is the field $\mathbb{Q}$, and such extensions of finite degree of the field $\mathbb{Q}$ are the fields used in quantum computing, we shall restrict to this particular case.

**Proposition 11 (Exension of finite degree of the field $\mathbb{Q}$).** *If the field $\langle L, +, \times \rangle$ is an extension of finite degree of the field $\langle \mathbb{Q}, +, \times \rangle$, then computability is stable over $\langle L, +, \times \rangle$.*

*Proof.* We have proved in Proposition 10, the existence of an admissible indexing $s, i$ of the structure $\langle K, L, +, \times, +, ., \times \rangle$. The function $i$ is obviously an admissible indexing of the field $\langle L, +, \times \rangle$.

Now consider an arbitrary admissible indexing $j$ of this structure and let us prove that $j^{-1}(\mathbb{Q})$ is effectively enumerable. The set $dom(j)$ is recursively enumerable, thus let $g$ be a generating function, we define $\hat{-}$ as the function mapping $n$ and $p$ to $g(z)$ where $z$ is the least natural number such that $p \hat{+} g(z) = n$, and we define $\hat{/}$ in a similar way. Let $z$ be a natural number such that $j(z) = 0$ and $u$ be a natural number such that $j(u) = 1$. Let $J$ be the computable function from $\mathbb{N}$ to $dom(j)$ defined by $J(0) = z$ and $J(p+1) = J(p) \hat{+} u$. Let $f$ be the computable function, from $\mathbb{N}^3$ to $\mathbb{N}$, mapping $p$, $q$ and $r$ to $(J(p) \hat{-} J(q)) \hat{/} (J(r) \hat{+} u)$. The set $j^{-1}(\mathbb{Q})$ is the image of this function. Thus, it is effectively enumerable.

Let $j$ and $j'$ be two indexings of $\langle L, +, \times \rangle$. The functions $j_{|j^{-1}(\mathbb{Q})}, j$ are an indexing of $\langle \mathbb{Q}, L, +, \times, +, ., \times \rangle$ and $j'_{|j'^{-1}(\mathbb{Q})}, j'$ also. Thus, using Proposition 10, there exists a computable function $k$ such that $j' = j \circ k$.

## 6 Tensor spaces

In Proposition 9, we have shown that infinite-dimensional vector spaces do not have a stable notion of computability. Intuitively, the lack stable computability for infinite-dimensional vector spaces happens for the same reason as it does for $\langle \mathbb{N} \rangle$, as can be seen from the proofs of Proposition 8 and 9. This is because neither algebraic structures are effectively generated, i.e. there is an infinity of elements (the natural numbers in one case, the vectors of a basis in the other) which are unrelated from one another.

In the case of the natural numbers this can be fixed by requiring that the successor be computable, i.e. by considering the structure $\langle \mathbb{N}, S \rangle$. On the set of finite sequences of elements taken in a finite set, the problem would be fixed in a similar way by adding the *cons* operation, that adds an element at the head of the list. On an infinite set of trees, we would add the operation that builds the tree $f(t_1, \ldots, t_n)$ from $f$ and $t_1, \ldots, t_n, \ldots$ These operations express that, although these data types are infinite, they are finitely generated.

In the same way, in quantum computing, an infinite data type comes with some structure, *i.e.* the vector space used to describe these data has a basis that is finitely generated with the tensor product. Typically, the basis vectors have the form $b_1 \otimes \ldots \otimes b_n$ where each $b_i$ is either $|0\rangle$ or $|1\rangle$.

**Definition 8 (Tensor space).** *A tensor space $\langle K, E, +, \times, +, ., \otimes \rangle$ is a vector space with an extra operation $\otimes$ that is a bilinear function from $E \times E$ to $E$.*

**Theorem 2.** *Let $\langle K, E, +, \times, +, ., \otimes \rangle$ be a tensor space such that computability is stable over the field $\langle K, +, \times \rangle$ and there exists a finite subset $A$ of $E$ such that the set of vectors of the form $b_1 \otimes (b_2 \otimes \ldots \otimes (b_{n-1} \otimes b_n) \ldots)$, for $b_1, \ldots, b_n$ in $A$, is a basis of $E$. Then, computability is stable over the structure $\langle K, E, +, \times, +, ., \otimes \rangle$.*

*Proof.* Let $s$ be an indexing of the field $\langle K, +, \times \rangle$. Consider the terms of the form $(S^{\lambda_0}(0).(b_1^0 \otimes \ldots \otimes b_{n_0}^0) + (S^{\lambda_1}(0).(b_1^1 \otimes \ldots \otimes b_{n_1}^1) + \ldots$ where $\lambda_0, \lambda_1, \ldots$ are in $dom(s)$ and $b_1^0, b_2^0, \ldots$ are in $A$. The denotation of such a term is the vector $s(\lambda_0).(b_1^0 \otimes \ldots \otimes b_{n_0}^0) + s(\lambda_1).(b_1^1 \otimes \ldots \otimes b_{n_1}^1) + \ldots$ As $s$ is surjective and the vectors $b_1^0 \otimes \ldots \otimes b_{n_0}^0, \ldots$ form a basis, every element of $E$ is the denotation of a term. As equality of denotations can be decided on scalars, it can be decided on terms. Thus, the structure $\langle E, +, ., \otimes \rangle$ is finitely generated relatively to the field $\langle K, +, \times \rangle$. As computability is stable over this field, it is stable over $\langle K, E, +, \times, +, ., \otimes \rangle$.

## 7 Conclusion

The robustness of the notion of computability over the natural numbers has been pointed out for long. But it has also been pointed out that this robustness does not extend to other countable domains, where computability is relative to the choice of an indexing. We have shown that this robustness is, in fact, shared by many algebraic structures: all the extensions of finite degree of the field of rationals (*e.g.* $\mathbb{Q}[\sqrt{2}]$ and $\mathbb{Q}[i, \sqrt{2}]$), all the finite-dimensional vector spaces over such a field, and all tensor spaces over such a field that are finite-dimensional (as tensor spaces) even if they are infinite-dimensional (as vector spaces).

For the vector spaces used in quantum computing, it is not the dimension as a vector space that matters, but the dimension as a tensor space. Indeed, the vector space operations handle the superposition principle, but the finite generation of the data types is handled by the tensor product. Finite-dimensional tensor space over extensions of finite degree of the field of rationals are probably sufficient to express all quantum algorithms.

Whether such spaces are sufficient to express all quantum physics is related to the possibility to decompose a system that has an infinite number of base states into finite but *a priori* unbounded number of subsystems, each having a finite number of base states.

More generally, when a structure $E$ has stable computability, and some dynamics of a physical system is described as a function $f$ $E$ to $E$, we can then consider its associated function $\hat{f}$, and ask ourselves whether this function is computable, universal, or uncomputable. Hence, this provides a formal sense in which some dynamics may respect, or break the Church-Turing thesis [11].